\documentclass[twocolumn,pra,aps,superscriptaddress]{revtex4}

\usepackage{epsfig}
\usepackage{graphicx}
\usepackage{dcolumn}
\usepackage{bm} 
\usepackage{epstopdf}
\usepackage{amsmath}

\newcommand{\beq}{\begin{equation}}
\newcommand{\eeq}{\end{equation}}
\newcommand{\beqa}{\begin{eqnarray}}
\newcommand{\eeqa}{\end{eqnarray}}

\def\oc#1{{ Opt.\ Commun.} {\bf#1}}
\def\jmo#1{{ J.\ Mod.\ Opt.} {\bf#1}}
\def\jpb#1{{ J.\ Phys.\ B} {\bf#1}}

\def\pra#1{{ Phys.\ Rev. A\/} {\bf#1}}

\def\prl#1{{ Phys.\ Rev.\ Lett.} {\bf#1}}

\def\rmp#1{{ Rev.\ Mod.\ Phys.} {\bf#1}}
\def\sci#1{{ Science} {\bf#1}}
\def\nat#1{{ Nature} {\bf#1}}

\begin{document}

\title{On the longitudinal momentum of the electron at the tunneling exit}

\author{Ruihua Xu}
\affiliation{Graduate School, China Academy of Engineering Physics, Beijing 100193, China}

\author{Tao Li}
\affiliation{Beijing Computational Science Research Center, Beijing 100193, China}

\author{Xu Wang}
\email{xwang@gscaep.ac.cn}
\affiliation{Graduate School, China Academy of Engineering Physics, Beijing 100193, China}

\date{\today}

\begin{abstract} The longitudinal momentum of the electron at the tunneling exit is a useful quantity to make sense of the tunneling ionization process. It was usually assumed to be zero from a classical argument, but recent experiments show that it must be nonzero in order to explain the measured electron momentum distributions. In this article we show that the flow momentum of the probability fluid is a sensible quantum mechanical definition for tunneling-exit momentum, and it can be (and in general is) nonzero at the tunneling exit point where the kinetic energy is zero by definition. We show that this longitudinal momentum is nonzero even in the static or adiabatic limit, and this nonzero momentum is a purely quantum mechanical effect determined by the shape of the wave function in the vicinity of the tunneling exit point. Nonadiabaticity or finite wavelength may increase this momentum substantially, and the detailed value depends on both the atomic and the laser parameters.
\end{abstract}


\maketitle

\section{Introduction}

Tunneling ionization is widely accepted \cite{Corkum-93, Kulander-93} as the first step of many strong-field processes, including high harmonic generation \cite{McPherson-87, Ferray-88} and attosecond pulse generation \cite{Krausz-RMP-09}, nonsequential double and multiple ionization \cite{Walker-94,Palaniyappan-05,Eberly-RMP-12}, laser-induced electron diffraction \cite{Blaga-12, Wolter-16}, etc. A thorough understanding of tunneling ionization is thus beneficial for the understanding of these strong-field phenomena.

One understands the tunneling ionization process through specifying some directly related characteristics, for example, the rate of tunneling, the entrance and the exit position of tunneling, the time needed for tunneling, the electron momentum at the exit of tunneling, etc. Although some concepts (for instance, the tunneling time \cite{Landauer-Martin, Landsman-15}) are controversial and still under debate, active efforts for understanding and characterizing tunneling ionization are ongoing, both theoretically \cite{Czirjak-00, Ivanov-05, Teeny-16-PRL, Teeny-16-PRA, Ni-16, Ni-18, Tian-17, Ivanov-17, Zhang-17, Gao-17, Liu-17, Wang-18, Ni-18b} and experimentally \cite{Comtois-05, Eckle-08, Arissian-10, Pfeiffer-12, Boge-13, Fechner-14, Sun-14, Camus-17, Han-17}.

The focus of the current article will be on one of the tunneling characteristics, namely, the electron momentum at the tunneling exit, or the longitudinal momentum parallel to the laser polarization direction in two- or three-dimensional tunneling problems. (The transverse momentum perpendicular to the laser polarization direction is relatively simple \cite{Ivanov-05, Arissian-10} and will not be the focus of the current article.) At first glance from a classical perspective, this tunneling-exit momentum should not be a problem at all: At the tunneling exit point the kinetic energy of the electron is zero (since the potential energy equals to the total energy) therefore the momentum should be zero. Indeed, several versions of the semiclassical trajectory models \cite{Brabec-96, Fu-01, Yudin-01, Cai-17, Li-17} still assume the longitudinal momentum to be zero based on this classical argument.

However, recent experiments have found that assuming the longitudinal momentum to be zero does not yield good agreements between theoretically simulated and experimentally measured electron momentum distributions \cite{Comtois-05, Pfeiffer-12, Camus-17}. The simulations are usually done using two-step semiclassical trajectory methods: The electron is first emitted via quantum tunneling with assumed tunneling-exit characteristics including tunneling-exit position and momentum, then the electron is treated as a classical particle traveling in the combined Coulomb and laser field. Only when nonzero (usually a fraction of an atomic unit) longitudinal momenta at the tunneling exit are included do the simulated electron momentum distributions agree well with the measurements.

Then the following questions arise from the theoretical point of view: In quantum mechanics what is the definition of electron momentum at specified positions (for example, the tunneling exit)? That is, what momentum are we exactly talking about quantum mechanically? How can the electron momentum be nonzero at a point with zero kinetic energy, as discovered by the above-cited experiments? What determines the value of the longitudinal momentum at the tunneling exit?

The goal of the current article is to answer these questions. We advocate the usage of the flow momentum of the probability fluid as the definition of electron momentum at specified positions. The flow momentum is a widely used concept in the field of quantum hydrodynamics \cite{Wyatt-book}, which views the time-dependent evolution of the wave function as a probability flow in real space. Nonzero flow momentum is allowed at the tunneling exit point where the kinetic energy is zero. The justification of using the flow momentum will be clearly demonstrated in the textbook example of tunneling through a one-dimensional (1D) square potential barrier. The nonzero momentum at the tunneling exit point is shown to be a purely quantum mechanical effect without a classical correspondence, therefore arguments based on classical mechanics cannot explain the existence of this nonzero momentum. The nonzero tunneling momentum is determined by the shape of the wave function in the vicinity of the tunneling exit point. Nonadiabaticity or finite wavelength may increase this momentum substantially, and the detailed value depends on both the atomic and the laser parameters.

This article is organized as follows. In Section II we introduce the flow momentum and its behavior in the static/adiabatic limit. Then we demonstrate the justification of using the flow momentum in the textbook example of tunneling through a 1D square potential barrier. Numerical methods of solving the (time-independent or time-dependent) Schr\"odinger equations will also be introduced. In Section III we present numerical results and discussions on the longitudinal momentum at the tunneling exit. A summary will be given in Section IV.

\section{Method}

\subsection{The flow momentum of the probability fluid}

The time evolution of a wave function can be viewed as a dynamic flow of the probability fluid in the real space. Based on this perspective the research area of quantum hydrodynamics is developed \cite{Wyatt-book}. We can write a wave function $\Psi(x,t)$ in the following polar form
\beq
\Psi(x,t) = A(x,t) e^{i\phi(x,t)},
\eeq
with $A(x,t) \ge 0$ the amplitude function and  $\phi(x,t)$ the phase function, both of which are real.

The probability current (or flux) can be obtained (atomic units are used)
\beqa
j (x,t) &=& \frac{i}{2} \left[ \Psi(x,t) \frac{\partial}{\partial x} \Psi^*(x,t) -\text{c.c.} \right]    \\
&=& \rho(x,t) \frac{\partial \phi(x,t)}{\partial x}
\eeqa
where $\rho (x,t) = |\Psi(x,t)|^2 = A^2(x,t)$ is the probability density. In analogy to the classical fluid equation $j (x,t) = \rho (x,t) v(x,t)$, one sees that the flow velocity or momentum of the quantum mechanical probability fluid is the spatial derivative of the phase function
\beq
p (x,t) =   \frac{\partial \phi(x,t)}{\partial x}.    \label{e.flowmtm}
\eeq
We emphasize that this flow momentum $p(x,t)$ is legally defined on each position $x$. The flow momentum has also been used in strong-field atomic physics and it is sometimes called the ``virtual-detector" momentum \cite{Thumm-03, Wang-13, Teeny-16-PRL, Teeny-16-PRA, Ni-16, Ni-18, Tian-17, Wang-18}, imagining putting a virtual detector at position $x$ and extracting the flow momentum at that point.

\subsection{Flow momentum in the static/adiabatic limit}

In the static or the adiabatic limit the time-independent Schr\"odinger equation (TISE) reads
\beq
-\frac{1}{2} \psi''(x) + V(x) \psi(x) = E \psi(x),  \label{e.TISE}
\eeq
where the potential $V(x)$ is time independent and the double prime denotes the second derivative in space. The TISE can be rearranged as
\beqa
\psi''(x) &=& -2 [E-V(x)] \psi(x) \nonumber \\
           &=& - p_\text{cl}^2(x) \psi(x),    \label{e.TISE2}
\eeqa
where $p_\text{cl}(x) \equiv \sqrt{2[E-V(x)]}$ is the same as the classical momentum for $E\ge V(x)$. In the classical forbidden region where $E<V(x)$, $p_\text{cl}(x)$ becomes purely imaginary. At turning points, for example a tunneling exit point $x_e$, $E=V(x_e)$, so $p_\text{cl}(x_e) = 0$. This is the argument based on which the tunneling exit momentum was assumed to be zero in semiclassical trajectory models \cite{Brabec-96, Fu-01, Yudin-01, Cai-17, Li-17}.

Substituting $\psi(x) = A(x) e^{i\phi(x)}$ into Eq. (\ref{e.TISE2}) and collecting separately the real part and the imaginary part, one gets the following two equations that are equivalent to the original TISE
\beqa
& & A'' - A(\phi')^2 + p_{cl}^2 A = 0,   \label{e.TISE_Re} \\
& & A \phi'' + 2 A' \phi' = 0, \ \ \text{or} \ \ (A^2 \phi')'=0.  \label{e.TISE_Im}
\eeqa
Eq. (\ref{e.TISE_Im}) tells that $A^2 \phi' = A^2 p = C$, or $p(x) = C/A^2(x)$, i.e., the flow momentum is larger where the probability density is smaller, vice versa.

We may rewrite Eq. (\ref{e.TISE_Re}) by dividing $A(x)$ on both sides
\beq \label{e.flowmtmbreak}
p^2(x) \equiv [\phi'(x)]^2 = p_\text{cl}^2(x) + \frac{A''(x)}{A(x)}.
\eeq
One sees that the flow momentum $p(x)$ has two contributing terms: one is the classical momentum $p_\text{cl}(x)$, and the other is the $A''/A$ term. At the tunneling exit point $x_{e}$, $p_\text{cl}(x_e) = 0$, then the flow momentum is
\beq
p(x_e) = \sqrt{ \frac{A''(x_e)}{A(x_e)} }.
\eeq
The flow momentum is nonzero if $A''/A$ is nonzero at the tunneling exit point. 

Having the unit of energy, the $A''/A$ term is also called the ``shape kinetic energy" \cite{Wyatt-book}, since it is determined by the shape of the wave function (the second derivative of $A(x)$ normalized by its absolute value). This shape kinetic energy is a purely quantum mechanical effect without a classical correspondence. Arguments based on classical mechanics necessarily miss this energy.

\subsection{Tunneling through a 1D square potential barrier}

\begin{figure}
  \centering
  \includegraphics[width=4.2cm,height=3.5cm]{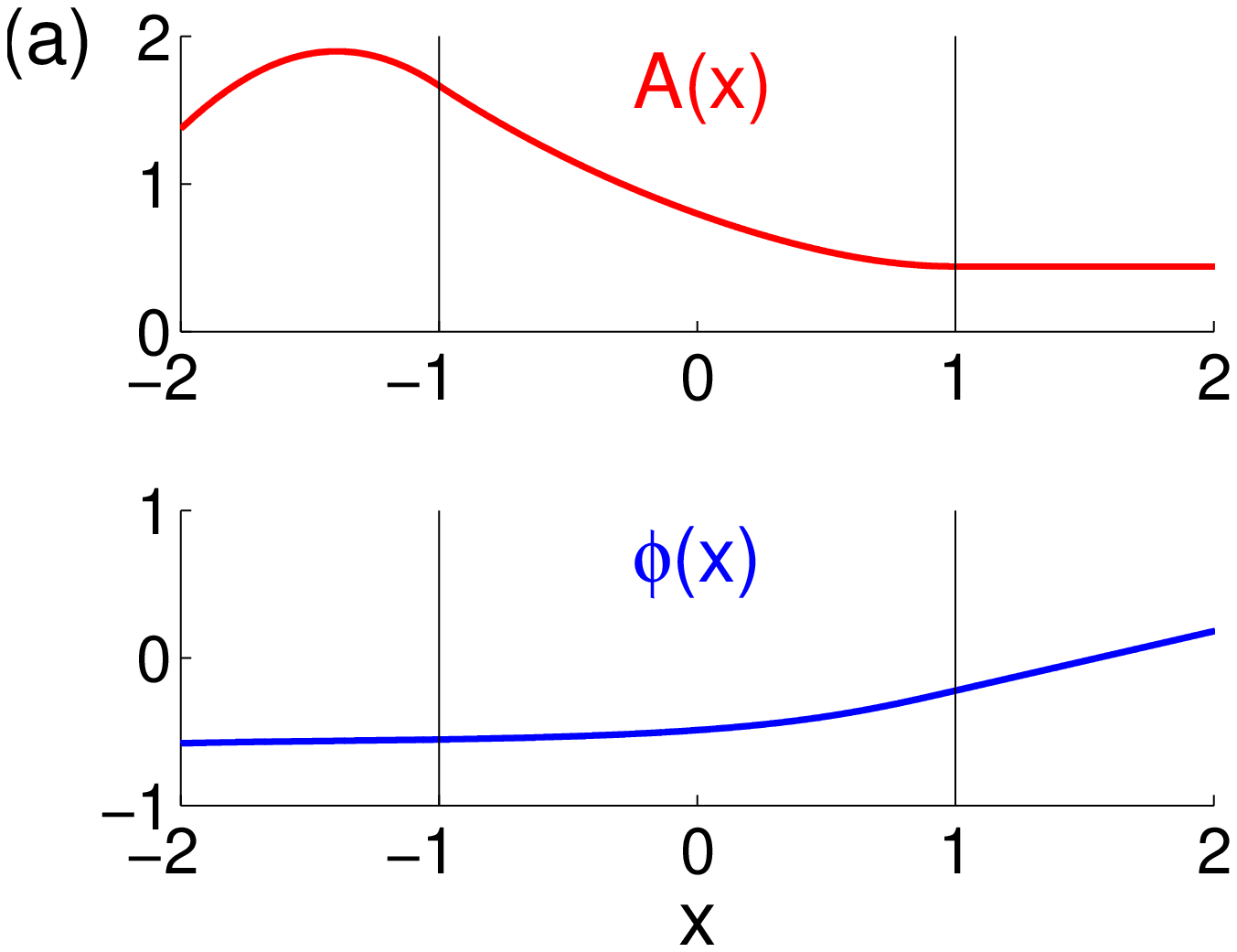}
  \includegraphics[width=4.2cm,height=3.5cm]{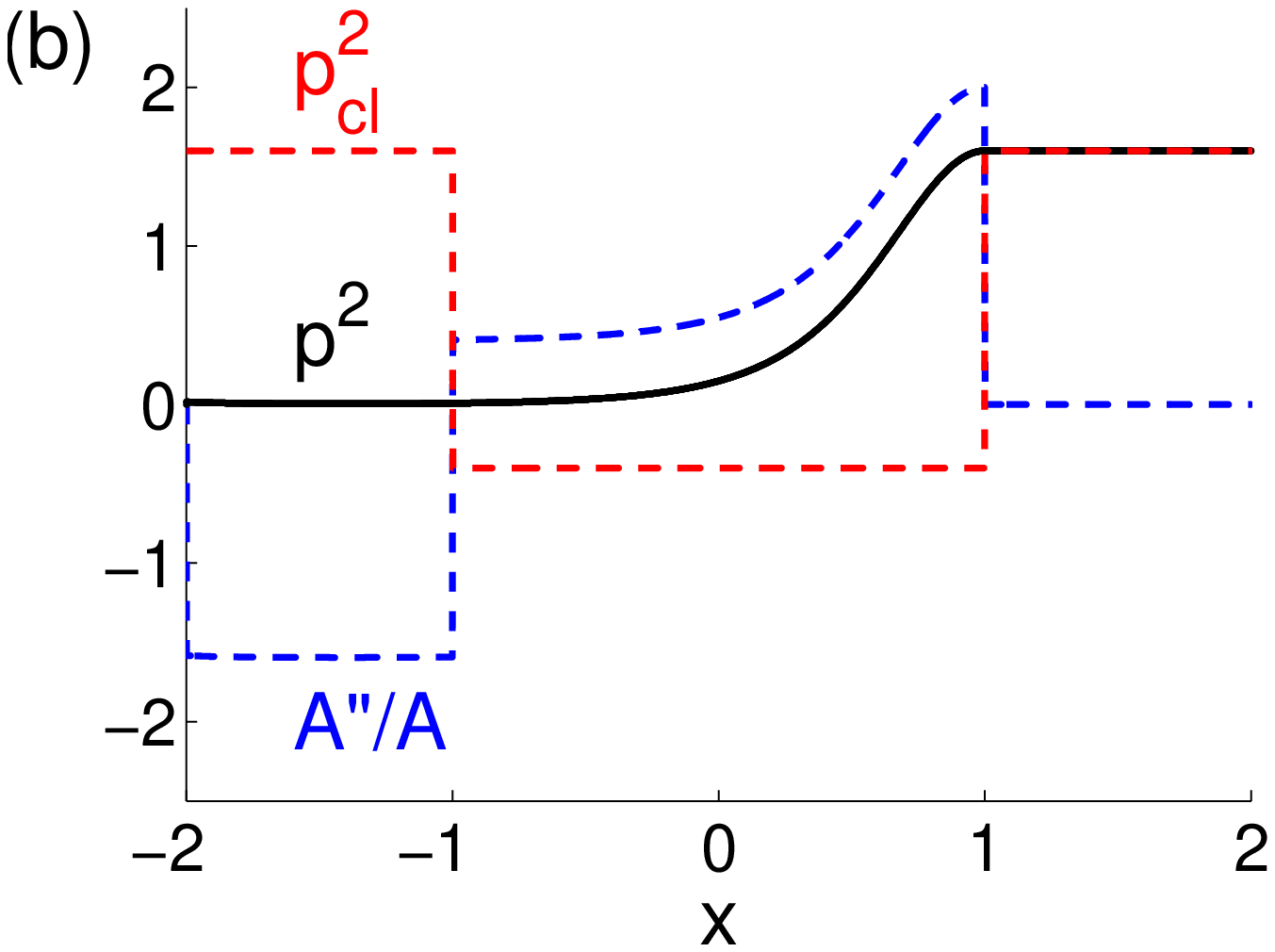}
  \caption{(a) The amplitude $A(x)$ and the phase $\phi(x)$ of the wave function $\psi(x)$ in Eq. (\ref{e.psi_squarebarrier}). The parameter values used are $a=1$, $V_0=1$, and $E=0.8$. The vertical lines show the boundaries of the potential barrier. (b) $p^2(x)$ (black solid), $p_\text{cl}^2(x)$ (red dashed), and $A''(x)/A(x)$ (blue dashed) for this potential barrier. The first term equals to the sum of the last two terms from Eq. (\ref{e.flowmtmbreak}).}\label{f.square}
\end{figure}

In this subsection we use the textbook example of tunneling through a 1D square barrier potential to illustrate the flow momentum $p(x)$, as well as its two contributing terms, namely, the semiclassical momentum $p_\text{cl}(x)$ and the momentum corresponding to the shape kinetic energy $A''(x)/A(x)$.

Consider a potential barrier $V(x) = V_0$ for $-a \le x \le a$ and zero otherwise. A plane wave $e^{ikx}$ comes from left and travels to the right, with energy $E = k^2/2 < V_0$. It is partially reflected at the boundary $x = -a$ and partially transmitted via quantum tunneling through the barrier to the region $x>a$. The wave function can be written piecewise as
\beq   \label{e.psi_squarebarrier}
\psi(x) = \left \{ \begin{array}{ll}
                    e^{ikx} + B e^{-ikx}, & \text{ if} \ x<-a \\
                    C e^{\kappa x} + D e^{-\kappa x}, & \text{ if} \ -a\le x<a \\
                    F e^{ikx}, & \text{ if} \ x\ge a
                  \end{array}  \right.
\eeq
where $B$, $C$, $D$, $F$ are complex coefficients that can be determined by connecting conditions (requiring $\psi(x)$ and $\psi'(x)$ to be continuous) at the two boundaries. The detailed results of these coefficients will not be shown explicitly here. $\kappa = \sqrt{2(V_0-E)}$ is the wave vector inside the potential barrier.

Fig. \ref{f.square} (a) shows the amplitude and phase of the wave function $\psi(x)$ around the potential barrier. The parameter values used are $a=1$, $V_0=1$, and $E=0.8$. Fig. \ref{f.square} (b) shows the three terms in Eq. (\ref{e.flowmtmbreak}), namely, the square of the flow momentum $p^2(x)$, the square of the semiclassical momentum $p_\text{cl}^2(x)$ (being negative inside the potential barrier), and the shape kinetic energy $A''(x)/A(x)$. The first term equals to the sum of the last two terms.

The following observations favor the usage of the flow momentum $p(x)$ over the semiclassical momentum $p_\text{cl}(x)$. First, $p(x)$ is always real (since it is the derivative of a real function $\phi(x)$) but $p_\text{cl}(x)$ can be real (in the classical allowed region) or imaginary (in the classical forbidden region). Second, $p(x)$ is continuous but $p_\text{cl}(x)$ may not. This can be seen from Fig. \ref{f.square} (b) where $p_\text{cl}^2(x)$ suffers discontinuity at the boundaries.

Starting from (and including) the tunneling exit point $x_e=a$, the wave function is a plane wave $\sim e^{ikx}$, so the electron has momentum $k$ at the tunneling exit point. This textbook example shows clearly that the electron momentum does not have to be zero at the tunneling exit point. The flow momentum $p(x)$ catches the nonzero feature but the semiclassical momentum $p_\text{cl}(x)$ does not.

\subsection{Numerically solving the TISE}

\begin{figure} [t!]
  \centering
  \includegraphics[width=8cm]{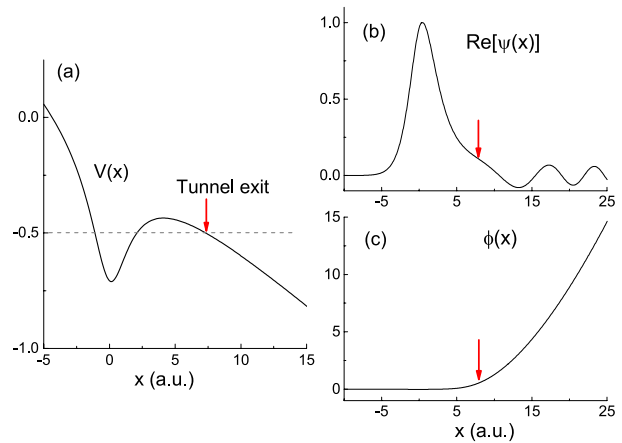}
  \caption{(a) A laser-tilted soft-core Coulomb potential. Without the laser field, the energy of the ground state is -0.5 a.u. The red arrow in each panel indicates the position of the tunneling exit point. (b) The real part of the wave function $\psi(x)$. (c) The phase of the wave function.}\label{f.1Dhydrogen}
\end{figure}

For a general potential, such as a tilted Coulomb potential, the wave function does not have analytical solutions and the TISE needs to be solved numerically. In this article we use the Numerov's method \cite{Numerov}, which is a widely used numerical method of solving the TISE.

Consider tunneling ionization from a laser-tilted soft-core Coulomb potential \cite{SoftCore1, SoftCore2} 
\beq
V(x)= -\frac{1}{\sqrt{x^2+2}}-F_0 x,
\eeq
as illustrated in Fig. \ref{f.1Dhydrogen} (a) with laser field strength $F_0 = 0.05$ a.u. This soft-core Coulomb potential has a ground state energy of -0.5 a.u. (-13.6 eV). (With the laser field, a small Stark shift appears but it will not affect our discussions or conclusions.) The tunneling exit point can be determined to be $x_e \approx 7.3$ a.u. The wave function can be solved numerically using Numerov's algorithm
\beq  \label{e.Numerov}
\psi_{n+1} = \frac{2(1-\frac{5}{12} h^2 k_n^2)\psi_n - (1+\frac{1}{12} h^2 k_{n-1}^2)\psi_{n-1}}{1+\frac{1}{12} h^2 k_{n+1}^2},
\eeq
where $x$ has been discretized as $x_n = x_0 + nh$ with $x_0$ the boundary on one side and $h$ the step size. $k_n = \sqrt{2[E - V(x_n)]}$ is the wave vector at $x_n$ and it is imaginary in the classical forbidden region.

Knowing $\psi_{n-1}$ and $\psi_n$, the next function value $\psi_{n+1}$ can be obtained using Eq. (\ref{e.Numerov}). This recursive process can be performed either way, from left to right or from right to left. In practice for stability considerations the integration is performed from both ends and the wave function is connected smoothly near the peak. Integrating from right to left, we impose the outgoing boundary condition
\beq
\psi(x\rightarrow \infty) \sim e^{ikx};
\eeq
And integrating from left to right, the boundary condition
\beq
\psi(x\rightarrow -\infty) \rightarrow 0
\eeq
is used. The connection is performed around the peak of the wave function where the first derivative vanishes.

The real part and the phase of the numerical wave function is shown in Fig. \ref{f.1Dhydrogen} (b) and (c), respectively. The red arrow on each panel shows the position of the tunneling exit. One immediately sees that the slope of the phase function $\phi(x)$ is nonzero at the tunneling exit, telling a nonzero tunneling-exit momentum.

\subsection{Numerically solving the TDSE}

The TDSE for an atom interacting with an external laser field can be written as
\begin{equation}
i\frac{\partial}{\partial t}\Psi(\vec{r},t)=\hat{H}\Psi(\vec{r},t)=[\hat{H}_{0}+\hat{H}_{I}]\Psi(\vec{r},t),
\end{equation}
where $\hat{H}_{0}$ is the field-free Hamiltonian and $\hat{H}_{I}$ is the atom-field interaction
\begin{eqnarray}
\hat{H}_{0} &=& -\frac{1}{2}\frac{d^{2}}{dr^2}+\frac{\hat{L}^2}{2r^2}+V(r),\\
\hat{H}_{I} &=& \vec{r} \cdot \vec{e}_{z} F(t) = F(t) r \cos \theta. \label{e.H_I}
\end{eqnarray}
For the hydrogen atom, $V(r)=-1/r$. For other atoms, a single-active-electron (SAE) approximation is used and the non-active ion core is modeled by the Green-Sellin-Zachor (GSZ) potential \cite{Green1969}
\begin{equation}
V(r)=-\frac{1}{r}\left[\frac{Z-1}{(\eta/\xi)(e^{\xi r}-1)+1}+1 \right].
\end{equation}
The values of $Z$, $\eta$, and $\xi$ for different noble gas atoms are listed in Table \ref{Tab1}.

Here we have used the length-gauge form of the interaction Hamiltonian. The laser field $F(t)= F_0\sin\omega t$ is assumed to be linearly polarized along the $z$ direction with amplitude $F_0$ and angular frequency $\omega$.

We use a generalized pseudospectral method \cite{Tong1997} to numerically solve the TDSE.
The Schr\"odinger equation can be propagated in discrete time steps as
\begin{eqnarray}
\Psi(\vec{r},t+\Delta t)&\simeq&\exp(-i\hat{H}_{0}\Delta t/2 )\nonumber\\
&\times& \exp[-i\hat{H}_{I}(r,\theta,t+\Delta t)\Delta t]\nonumber\\
&\times& \exp(-i\hat{H}_{0}\Delta t/2)\Psi(\vec{r},t).
\label{psiyanhua}
\end{eqnarray}
The time propagation of the wave function from $t$ to $t+\Delta t$ is achieved by three steps: (i) Propagation for half a time step $\Delta t/2$ in the energy space spanned by $\hat{H}_{0}$; (ii) Transformation to the coordinate space and propagation for one time step $\Delta t$ under the atom-field interaction $\hat{H}_{I}$; (iii) Transformation back to the energy space spanned by $\hat{H}_{0}$ and propagation for another half time step $\Delta t/2$. The commutation errors are on the order of $\Delta t^3$.

The wave function $\Psi(\vec{r},t)$ can be expanded in Legendre polynomials
\begin{eqnarray}
\Psi(r_{i},\theta_{j},t)=\sum_{l=0}^{lmax}g_{l}(r_{i})P_{l}(\cos\theta_{j}),
\end{eqnarray}
if the atom is initially in an $s$-state (the magnetic quantum number $m=0$) and the laser polarization is linear ($\Delta m=0$). The $g_{l}(r_{i})$ is calculated by the Gauss-Legendre quadrature
\begin{eqnarray}
g_{l}(r_{i})=\sum_{k=1}^{L+1}w_{k}P_{l}(\cos\theta_{k})\Psi(r_{i},\theta_{k},t),
\end{eqnarray}
where quadrature lattices ${\cos\theta_{k}}$ are zeros of the Legendre polynomials $P_{l+1}(\cos\theta_{k})$ and ${w_{k}}$ is the corresponding quadrature weight.

Now the evolution of the wave function in the energy space spanned by $\hat{H}_{0}$  can be written as
\begin{align}
\exp(&-i\hat{H}_{0}\Delta t/2)\Psi(r_{i},\theta_{j},t) \nonumber \\
&=\sum_{l=0}^{lmax}[\exp(-i\hat{H}_{0}^l\Delta t/2)g_{l}(r_{i},t)]P_{l}(\cos\theta_{j}).
\label{eq3.23}
\end{align}
Each $g_{l}$ is propagated independently within individual $\hat{H}_{0}^{l}$ energy space.

In order to avoid artificial boundary reflection, for each time step a mask function $M(r)=\cos^{1/4}[(r-r_{0})/(r_{m}-r_{0})\pi/2]$ is multiplied to the wave function for $r\ge r_0$. Here $r_{0}$ is the entrance radius of the absorbing region and $r_{m}$ is the radius of the numerical grid.

\begin{table}
\begin{tabular}{ccccc}
  \hline
   Atom & ~~$Z$       & ~~~~$I_p$ & $~~~~~\xi$ & $~~~~~\eta$ \\
  \hline He~~ &~~2   & ~~~~-0.9  &~~~~ $ 2.625 $  &~~~~ $1.770$  \\
   Ne~~  &~~10        & ~~~~-0.79  &~~~~ $1.792$  &~~~~ $2.710$  \\
   Ar~~  &~~18        & ~~~~-0.58  &~~~~ $0.957$  &~~~~ $3.500$ \\
   Kr~~  &~~36        & ~~~~-0.51 &~~~~ $1.351$  &~~~~ $4.418$ \\
   Xe~~  &~~54        & ~~~~-0.45 &~~~~ $1.044$  &~~~~ $5.101$ \\
  \hline
\end{tabular}
  \caption{Values of $Z$, $\eta$, and $\xi$ used in the GSZ model potential for noble gas atoms. From Ref. \cite{Green1969}.}\label{Tab1}
\end{table}

\section{Results and discussions}

\begin{figure}
  \centering
  \includegraphics[width=8.5cm]{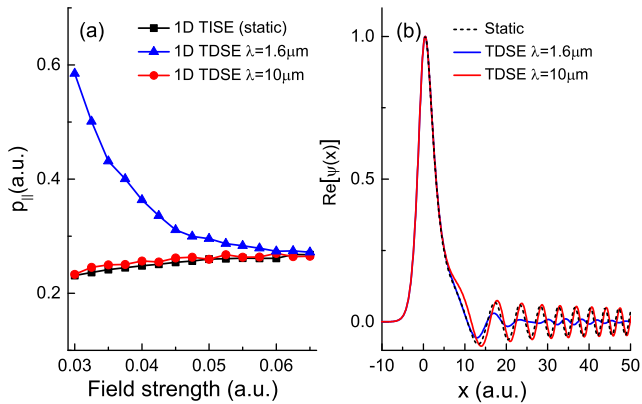}
  \caption{(a) Tunneling-exit momentum $p_{\parallel}$ as a function of laser field strength, for the 1D hydrogen atom. Black squares are for the static limit by solving the 1D TISE. Blue triangles are for $\lambda$ = 1.6 $\mu$m and red circles are for $\lambda$ = 10 $\mu$m, both by solving the 1D TDSE. See text for details. (b) Comparison of (the real part of) typical wave functions for the static case (black dashed curve), $\lambda$ = 1.6 $\mu$m (blue solid curve), and $\lambda$ = 10 $\mu$m (red solid curve). The laser field strength used is $F_0 = 0.05$ a.u.} \label{f.mtm_static}
\end{figure}

\subsection{Nonzero $p_{\parallel}$ in the static/adiabatic tunneling limit}

As explained previously in Section II (B), the time-independent Schr\"odinger equation tells that for a static tunneling problem, the longitudinal tunneling-exit momentum $p_{\parallel} \equiv p(x_e)$ is in general nonzero, unless for the situation $A''(x_e)=0$, i.e., the second derivative of the wave function amplitude vanishes at the tunneling exit $x_e$. This situation is rare. For tunneling ionization from Coulomb-like potentials,  $p_{\parallel}$ is in general nonzero in the static or the adiabatic limit.

This is indeed the case from numerical results. The curve with black squares in Fig. \ref{f.mtm_static} (a) shows $p_{\parallel}$ as a function of laser field strength for static tunneling ionization using the 1D soft-core hydrogen atom. The results are obtained by numerically solving the TISE using the Numerov's algorithm. One sees that $p_\parallel$ is nonzero for the entire range of field strengths used and takes values between 0.2 a.u. to 0.3 a.u. These momentum values are not small in strong-field atomic ionization and should not be simply neglected.

\subsection{Nonadiabatic effects on $p_\parallel$}

The static or adiabatic limit may be approached by using long wavelengths. If the wavelength is long, the laser electric field changes sufficiently slow such that at each time the electron experiences a static laser field. That is, the electronic wave function follows adiabatically the slowly changing laser field.

Indeed, TDSE calculations with $\lambda$ = 10 $\mu$m yield almost identical $p_\parallel$ values to the corresponding values in the static case, as shown in Fig. \ref{f.mtm_static} (a) (the curve with red circles). The almost perfect agreement for the $p_\parallel$ values indicates that the adiabatic approximation holds well for $\lambda$ = 10 $\mu$m, consistent to general expectations.

For shorter wavelengths, the adiabatic approximation may break down: The changing of the laser field may be too fast for the electronic wave function to follow adiabatically. This nonadiabaticity is expected to affect the values of $p_\parallel$. Indeed, for $\lambda$ = 1.6 $\mu$m, the values of $p_\parallel$ differ substantially from the corresponding static values, especially for relatively low field strengths, as shown in Fig. \ref{f.mtm_static} (a). At $F_0 = 0.03$ a.u., $p_\parallel$ for 1.6 $\mu$m can be as large as 0.6 a.u., more than twice the corresponding static or long-wavelength value. The difference decreases as the field strength increases though.

A comparison of wave functions is shown in Fig. \ref{f.mtm_static} (b), for the static, 10 $\mu$m, and 1.6 $\mu$m cases. The field strength used is $F_0 = 0.05$ a.u. One sees that the wave function for 10 $\mu$m agrees very well with the static case, whereas the wave function for 1.6 $\mu$m is visually different, with weaker amplitudes in the outside ionization region. 

To compare TDSE results to TISE results with field strength $F_0$, we set the laser field to have the form $F(t) = F_0 \sin\omega t$ and integrate TDSE for a quarter cycle from $t_i=0$ to $t_f=T/4=\pi/2\omega$. Then the TDSE wave function at $t_f$ is compared to the TISE wave function. The laser field strengths are the same for both cases then. Stopping the TDSE propagation at $T/4$ allows one to focus on direct tunneling ionization and avoid complications from recollision-related processes.

\subsection{Nonadiabatic transitions}

In the adiabatic limit, if the electron is initially in the ground state of the atom, as the laser field is turned on slowly, the electron is expected to stay in the ground state of the combined laser-atom system, although the state is no longer a bound state (It becomes a ``half-open" quasi-bound state as illustrated in Fig. \ref{f.1Dhydrogen}). Transition to other states of the laser-atom system is negligible.

Otherwise under conditions when the adiabatic approximation does not hold, transitions to other states may not be neglected. Here we calculate the probability of nonadiabatic transitions by projecting out the ground quasi-bound state from the numerical TDSE wave function 
\beq
P_\text{trans} = 1 - \left| \langle \psi_0(x) | \Psi(x,t_f) \rangle \right|^2,
\eeq
where $\psi_0(x)$ is the ground quasi-bound state, an example of which is illustrated in Fig. \ref{f.1Dhydrogen}. $\psi_0(x)$ is obtained by numerically solving the TISE, as explained in the previous section. $\Psi(x,t_f)$ is the numerical solution of the TDSE at time $t_f$. In the adiabatic limit $P_\text{trans}$ approaches zero, otherwise it will be greater than zero.

\begin{figure} [t!]
  \centering
  \includegraphics[width=8.5cm]{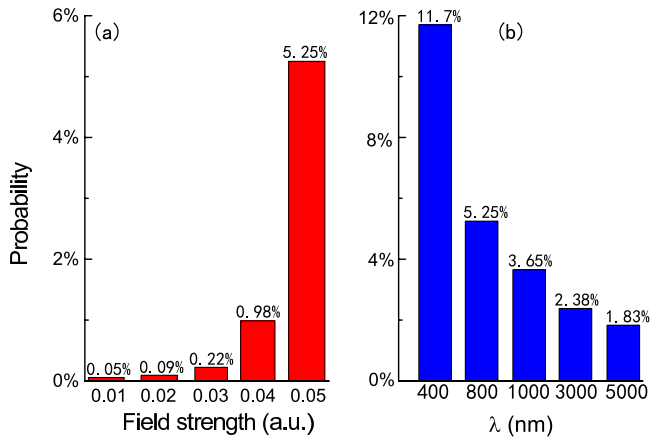}
  \caption{Probability of nonadiabatic transition (a) for different laser field strengths with the same wavelength of 0.8 $\mu$m, and (b) for different laser wavelengths with the same field strength of 0.05 a.u.}\label{f.nonadiabatic}
\end{figure}

Figure \ref{f.nonadiabatic} shows $P_\text{trans}$ (a) for different laser field strengths, under
the same laser wavelength of 0.8 $\mu$m; and (b) for different laser wavelengths, under the same laser field strength of 0.05 a.u. Under the same wavelength, $P_\text{trans}$ tends to be higher for higher field strengths due to stronger electron-laser coupling. And under the same field strength, $P_\text{trans}$ is higher for shorter wavelengths, as would be expected.

\subsection{3D results with different model atoms}

Up to now the results and analyses are obtained using the 1D soft-core hydrogen model atom. Simple as it is, the model has provided us with the following understandings: First, the longitudinal tunneling-exit momentum $p_\parallel$ is nonzero even in the static or the adiabatic limit; Second, nonadiabatic effects due to finite wavelengths may change $p_\parallel$ substantially.

\begin{figure} [t!]
  \centering
  \includegraphics[width=8.5cm,height=6.5cm]{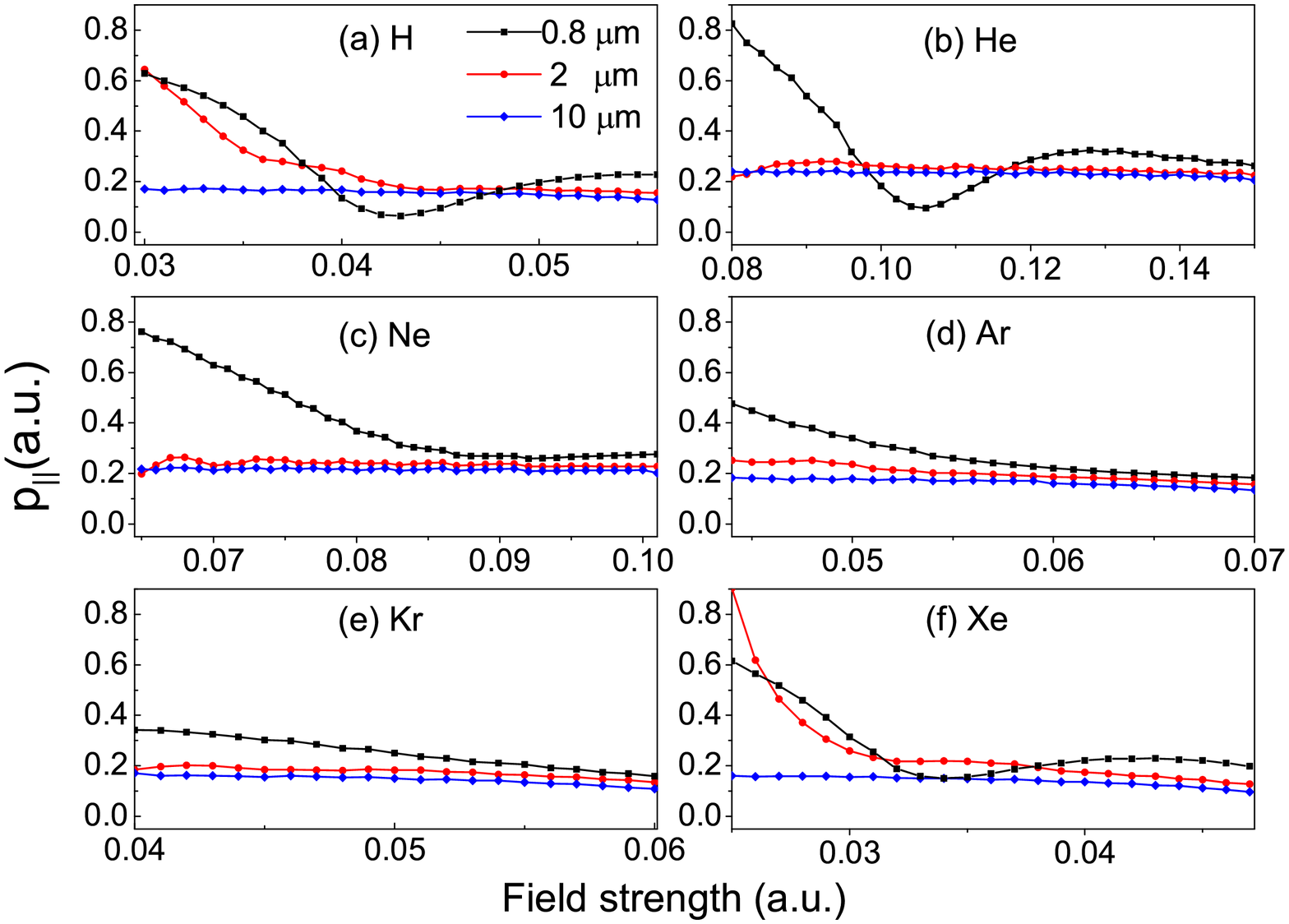}
  \caption{Longitudinal tunneling-exit momentum $p_\parallel$ for different model atoms, as labeled on each panel, obtained by numerically solving the 3D TDSE using the SAE approximation and the GSZ model potentials. Three wavelengths are used for each atom, namely, 0.8 $\mu$m (black squares), 2 $\mu$m (red circles), and 10 $\mu$m (blue dimands). }\label{f.3Dmtm}
\end{figure}

These conclusions also hold for 3D, though the detailed values of $p_\parallel$ may be slightly different. Fig. \ref{f.3Dmtm} shows 3D results of $p_\parallel$ for several model atoms including H, He, Ne, Ar, Kr, and Xe. The calculations are done by numerically solving the 3D TDSE using the GSZ model potentials \cite{Green1969}. The laser polarization is along the $z$-axis and $p_\parallel$ is also extracted from the wave function values along the $z$-axis. For simplicity, tunneling from points off the $z$-axis will not be discussed here. Due to cylindrical symmetry, the transverse momentum is zero for on-axis tunneling. 

Three wavelengths are used for each model atom, namely, 0.8 $\mu$m, 2 $\mu$m, and 10 $\mu$m. Let us first look at the results for 10 $\mu$m (blue diamonds), which are expected to approach the adiabatic limit. We see that with this wavelength, $p_\parallel$ is not very sensitive to atomic targets and it takes similar values (all around 0.2 a.u.) for different atoms. For the H atom, the 3D values are a little bit smaller than the corresponding 1D values, as shown in Fig. (\ref{f.mtm_static}). Also, in this adiabatic limit,  $p_\parallel$ is not very sensitive to the laser field strength either. 

Shorter wavelengths lead to complications on $p_\parallel$ due to nonadiabatic effects. For 2 $\mu$m, the values of $p_\parallel$ are close to the 10 $\mu$m values for some atoms such as He, Ne, Ar, and Kr, but H and Xe show exceptions. For 0.8 $\mu$m, the values of $p_\parallel$ are very different from the 10 $\mu$m values for all the six model atoms used. For shorter wavelengths, larger discrepancy appears with lower field strengths, and the general trend is that $p_\parallel$ decreases and approaches the long-wavelength values as the field strength increases. However, oscillations also appear especially with 0.8 $\mu$m, giving smaller $p_\parallel$ values than the corresponding long-wavelength ones with some field strengths.

\section{Summary}

In this article we consider the longitudinal momentum of the electron at the tunneling exit, one of the characteristics that help to understand tunneling ionization. Since the tunneling exit is defined as the point where the kinetic energy vanishes (the potential energy equals the total energy), at first glance from the classical perspective, one may simply think that the electron momentum is zero. Some versions of the semiclassical model still assume this zero-longitudinal-momentum condition \cite{Brabec-96, Fu-01, Yudin-01, Cai-17, Li-17}.

What brought the longitudinal momentum back into question was experiments \cite{Comtois-05, Pfeiffer-12, Camus-17}. Done with different schemes, they all reach the same conclusion that the longitudinal tunneling-exit momentum should be nonzero in order to explain the experimental data. Theoretical questions immediately arise: What momentum are we talking about in quantum mechanics that has a legal definition at a specified position? How can the momentum be nonzero at a point where the kinetic energy is zero? What determines the value of this momentum?

We advocate the usage of the flow momentum of the probability fluid and this flow momentum is legally defined at specified positions. The flow momentum has been a widely used concept in other research fields \cite{Wyatt-book} and it has also been used in strong-field atomic physics \cite{Thumm-03, Wang-13, Teeny-16-PRL, Teeny-16-PRA, Ni-16, Ni-18, Tian-17, Wang-18}. The flow momentum can be, and in general is, nonzero at the tunneling exit where the kinetic energy is zero. 

We show that the longitudinal momentum at the tunneling exit is nonzero even in the static or the adiabatic limit. This nonzero momentum is explained to be a purely quantum mechanical effect without a classical correspondence. It is determined by the shape of the wave function in the vicinity of the tunneling exit point. Values around 0.2 a.u. are predicted by numerical TDSE calculations in the long-wavelength limit.

We further show that nonadiabatic effects due to finite wavelengths can change the tunneling-exit momentum substantially. The values can be several times larger than the values in the adiabatic limit, especially for relatively low laser field strengths.

Acknowledgement: We acknowledge support from Science Challenge Project of China No. TZ2018005, National Key R\&D Program of China 2017YFA0403200, National Science Foundation of China No. 11774323, and NSAF No. U1730449.


\begin{thebibliography}{99}

\bibitem{Corkum-93} P. B. Corkum, \prl{71}, 1994 (1993).

\bibitem{Kulander-93} K. C. Kulander, K. J. Schafer, and J. L. Krause, in {\it Super-Intense Laser-Atom Physics}, edited by B. Piraux, A. L' Huillier, and K. Rzazewski (Plenum, New York,1993).

\bibitem{McPherson-87} A. McPherson, G. Gibson, H. Jara, U. Johann, T. S. Luk, I. A. McIntyre, K. Boyer, and C. K. Rhodes, JOSA B {\bf 4}, 595 (1987).

\bibitem{Ferray-88} M. Ferray, A. L'Huillier, X. F. Li, L. A. Lompre, G. Mainfray, and C. Manus, \jpb{21}, L31 (1988).

\bibitem{Krausz-RMP-09} F. Krausz and M. Ivanov, \rmp{81}, 163 (2009).

\bibitem{Walker-94} B. Walker, B. Sheehy, L. F. DiMauro, P. Agostini, K. J. Schafer, and K. C. Kulander, \prl{73}, 1227 (1994).

\bibitem{Palaniyappan-05} S. Palaniyappan, A. DiChiara, E. Chowdhury, A. Falkowski, G. Ongadi, E. L. Huskins, and B. C. Walker, \prl{94}, 243003 (2005).

\bibitem{Eberly-RMP-12} W. Becker, X. Liu, P. J. Ho, and J. H. Eberly, \rmp{84}, 1011 (2012).

\bibitem{Blaga-12}  C. I. Blaga, J. Xu,	A. D. DiChiara,	E. Sistrunk, K. Zhang, P. Agostini,	T. A. Miller,	L. F. DiMauro, and C. D. Lin, \nat{483}, 194 (2002).

\bibitem{Wolter-16} B. Wolter, M. G. Pullen, A.-T. Le, M. Baudisch, K. Doblhoff-Dier, A. Senftleben, M. Hemmer, C. D. Schr\"oter, J. Ullrich, T. Pfeiffer, R. Moshammer, S. Gr\"afe, O. Vendrell, C. D. Lin, and J. Biegert, \sci{354}, 308 (2016).

\bibitem{Landauer-Martin} R. Landauer and Th. Martin, \rmp{66}, 217 (1994).

\bibitem{Landsman-15} A. Landsman and U. Keller, Phys. Rep. {\bf 547}, 1 (2015).

\bibitem{Czirjak-00} A. Czirjak, R. Kopold, W. Becker, M. Kleber, and W. P. Schleich, \oc{179}, 29 (2000).

\bibitem{Ivanov-05} M. Yu Ivanov, M. Spanner, and O. Smirnova, \jmo{52}, 165 (2005).

\bibitem{Teeny-16-PRL} N. Teeny, E. Yakaboylu, H. Bauke, and C. H. Keitel, \prl{116}, 063003 (2016).

\bibitem{Teeny-16-PRA} N. Teeny, C. H. Keitel, and H. Bauke, \pra{94}, 022104 (2016).

\bibitem{Ni-16} H. Ni, U. Saalmann, and J. M. Rost, \prl{117}, 023002 (2016).

\bibitem{Ni-18} H. Ni, U. Saalmann, and J. M. Rost, \pra{97}, 013426 (2018).

\bibitem{Tian-17} J. Tian, X. Wang, and J. H. Eberly, \prl{118}, 213201 (2017).

\bibitem{Ivanov-17} I. A. Ivanov, C. H. Nam, and K. T. Kim, Sci. Rep. {\bf 7}, 39919 (2017).

\bibitem{Zhang-17} Q. Zhang, G. Basnayake, A. Winney, Y. Lin, D. Debrah, S. K. Lee, and W. Li, \pra{96}, 023422 (2017).

\bibitem{Gao-17} F. Gao, Y. Chen, G. Xin, J. Liu, and L. B. Fu, \pra{96}, 063414 (2017).

\bibitem{Liu-17} K. Liu and I. Barth, \prl{119}, 243204 (2017).

\bibitem{Wang-18} X. Wang, J. Tian, and J. H. Eberly, \jpb{51}, 084002 (2018).

\bibitem{Ni-18b} H. Ni, N. Eicke, C. Ruiz, J. Cai, F. Oppermann, N.I. Shvetsov-Shilovski, L.W. Pi, \pra{98}, 013411 (2018).

\bibitem{Comtois-05} D. Comtois, D. Zeidler, H. Pepin, J. C. Kieffer, D. M. Villeneuve, and P. B. Corkum, \jpb{38}, 1923 (2005).

\bibitem{Eckle-08} P. Eckle, A. N. Pfeiffer, C. Cirelli, A. Staudte, R.D\"orner, H. G. Muller, M. B\"utiker, and U. Keller, \sci{322}, 1525 (2008).

\bibitem{Arissian-10} L. Arissian, C. Smeenk, F. Turner, C. Trallero, A. V. Sokolov, D. M. Villeneuve, A. Staudte, and P. B. Corkum, \prl{105}, 133002 (2010).

\bibitem{Pfeiffer-12} A. N. Pfeiffer, C. Cirelli, A. S. Landsman, M. Smolarski, D. Dimitrovski, L. B. Madsen, and U. Keller, \prl{109}, 083002 (2012).

\bibitem{Boge-13} R. Boge, C. Cirelli, A. S. Landsman, S. Heuser, A. Ludwig, J. Maurer, M. Weger, L. Gallmann, and U. Keller, \prl{111}, 103003 (2013).

\bibitem{Fechner-14} L. Fechner, N. Camus, J. Ullrich, T. Pfeiffer, and R. Moshammer, \prl{112}, 213001 (2014).

\bibitem{Sun-14} X. Sun, M. Li, J. Yu, Y. Deng, Q. Gong, and Y. Liu, \pra{89}, 045402 (2014).

\bibitem{Camus-17} N. Camus, et al., \prl{119}, 023201 (2017).

\bibitem{Han-17} M. Han, M. Li, M. Liu, and Y. Liu, \pra{95}, 023406 (2017).

\bibitem{Brabec-96} T. Brabec, M. Y. Ivanov, and P. B. Corkum, \pra{54}, R2551 (1996).

\bibitem{Fu-01} L. B. Fu, J. Liu, J. Chen, and S. G. Chen, \pra{63}, 043416 (2001).

\bibitem{Yudin-01} G. L. Yudin and M. Y. Ivanov, \pra{63}, 033404 (2001).

\bibitem{Cai-17} J. Cai, Y. Chen, Q. Xia, D. Ye, and L. Fu, \pra{96}, 033413 (2017).

\bibitem{Li-17} X. Li, C. Wang, Z. Yuan, D. Ye, P. Ma, W. Hu, S. Luo, L. Fu, and D. Ding, \pra{96}, 033416 (2017).

\bibitem{Wyatt-book} R. E. Wyatt, {\it Quantum dynamics with trajectories: Introduction to quantum hydrodynamics}, Springer (2005).

\bibitem{Thumm-03} B. Feuerstein and U. Thumm, \jpb{36}, 707 (2003).

\bibitem{Wang-13} X. Wang, J. Tian, and J. H. Eberly, \prl{110}, 243001 (2013).

\bibitem{Numerov} B. V. Numerov, Astronomische Nachrichten {\bf 230}, 359 (1927).

\bibitem{SoftCore1} J. Javanainen, J. H. Eberly, and Q. Su, Phys. Rev. A 38, 3430 (1988).

\bibitem{SoftCore2} Q. Su and J. H. Eberly, Phys. Rev. A 44, 5997 (1991).

\bibitem{Green1969} A. E. S. Green, D. L. Sellin, and A. S. Zachor, Phys. Rev. {\bf184},1 (1969).

\bibitem{Tong1997} X.-M. Tong and S.-I Chu, Chem. Phys. {\bf 217}, 119 (1997).


\end{thebibliography}
\end{document}